\documentclass[notoc,hyper,letterpaper]{JHEP3}

\usepackage{epsfig}
\usepackage{amsbsy}
\usepackage{varioref}
\usepackage{pifont}
\newcommand{\dfrac}[2]{\frac{\strut \displaystyle{#1}}
                      {\strut \displaystyle{#2}}}

\title{Universal Non-Oblique Corrections\\
 in Higgsless Models and Beyond}

\author{R. Sekhar Chivukula and Elizabeth H. Simmons\\
Department of Physics and Astronomy, Michigan State University\\
East Lansing, MI 48824, USA\\
	E-mail: \email{sekhar@msu.edu, esimmons@msu.edu}}
	
\author{Hong-Jian He\\
Department of Physics, University of Texas\\
Austin, TX 78712, USA\\
	E-mail: \email{hjhe@physics.utexas.edu}}

\author{Masafumi Kurachi\\
Department of Physics, Nagoya University\\
Nagoya 464-8602, Japan\\
	E-mail:\email{kurachi@eken.phys.nagoya-u.ac.jp}}

\author{Masaharu Tanabashi\\
Department of Physics, Tohoku University\\
Sendai 980-8578, Japan\\
	E-mail:\email{tanabash@tuhep.phys.tohoku.ac.jp}}

\abstract{ Recently Barbieri, {\it et al.} have introduced a formalism to express the
deviations of electroweak interactions from their standard model forms
in ``universal'' theories, {\it i.e.} theories in which the corrections due to new physics 
can be expressed solely by modifications to the two-point correlation 
function of electroweak gauge currents of fermions. The parameters introduced by these 
authors are defined by the properties of the correlation functions at zero 
momentum, and differ from the quantities calculated by examining the
on-shell properties of the electroweak gauge bosons. In this letter we discuss the relationship
between the zero-momentum and on-shell parameters.  In addition, we present the
results of a calculation of these zero-momentum parameters
in an arbitrary Higgsless model in which the low-energy $\rho$ parameter is one and
which can be deconstructed to a linear chain of $SU(2)$ groups adjacent to a
chain of $U(1)$ groups. Our results demonstrate the importance of the 
universal ``non-oblique'' corrections which are present 
and elucidate the relationships among various calculations of electroweak
quantities in these models. Our expressions for these zero-momentum parameters depend
only on the spectrum of heavy vector-boson masses; therefore, the minimum size of the deviations
present in these models is related to the upper bound on the heavy vector-boson masses
derived from unitarity. We find that these models are disfavored by precision electroweak 
data, independent of any assumptions about the background metric or the behavior of the
bulk coupling.}

\preprint{ {MSUHEP-040824} \\
{DPNU-04-15} \\
{TU-728}}


\begin{document}

\section{Introduction}

The standard electroweak model is in excellent agreement with the majority of 
experimental data.  Despite this agreement,  the agent of 
electroweak symmetry breaking remains elusive. Furthermore the one-doublet Higgs model, 
the simplest way to accommodate symmetry breaking, is unsatisfactory. These observations
motivate the theoretical search for alternatives to the one-doublet Higgs model and the careful
examination of precision electroweak data to motivate or constrain extensions to the
standard model.

Recently, there have been interesting work on both of these fronts. 
On the theoretical side, ``Higgsless'' models of electroweak symmetry breaking have been proposed \cite{Csaki:2003dt}. 
Based on five-dimensional gauge theories compactified on an interval, these models 
achieve unitarity of electroweak boson self-interactions 
through the exchange of a tower of massive vector bosons 
\cite{SekharChivukula:2001hz,Chivukula:2002ej,Chivukula:2003kq}, 
rather than the exchange of a scalar Higgs boson \cite{Higgs:1964ia}.
Motivated by gauge/gravity duality \cite{Maldacena:1998re,Gubser:1998bc,Witten:1998qj,Aharony:1999ti}, models of this kind may be viewed as ``dual'' to more conventional models of dynamical symmetry breaking 
\cite{Weinberg:1979bn,Susskind:1979ms} such
as ``walking techicolor'' \cite{Holdom:1981rm,Holdom:1985sk,Yamawaki:1986zg,Appelquist:1986an,Appelquist:1987tr,Appelquist:1987fc}.

On the phenomenological side, Barbieri {\it et al.} \cite{Barbieri:2004qk}
have introduced a formalism to express the
deviations of electroweak interactions from their standard model forms 
in ``universal'' theories, {\it i.e.} theories in which the corrections due to new physics 
can be expressed solely by modifications to the two-point correlation 
function of electroweak gauge currents of fermions. The parameters introduced by these 
authors are defined by the properties of the correlation functions at zero 
momentum, and differ from the more familiar quantities calculated by examining the
on-shell properties of the electroweak gauge bosons.

In this letter we discuss the relationship
between the zero-momentum and on-shell parameters.  In addition, we present the
results of a calculation of these zero-momentum parameters
in a general class of Higgsless models in which the low-energy rho parameter is one and
which can be deconstructed \cite{Arkani-Hamed:2001ca,Hill:2000mu} 
to a linear chain of $SU(2)$ groups adjacent to a
chain of $U(1)$ groups. The details of the calculation of the zero-momentum
parameters in deconstructed higgsless models, which extend the results of  \cite{Chivukula:2004pk},  will be presented
in a forthcoming publication \cite{Chivukula2004}.

Our results demonstrate the importance of the 
universal ``non-oblique'' corrections which are present in these models 
and elucidate the relationships among various calculations of electroweak
quantities in these models
\cite{Csaki:2003zu,Nomura:2003du,Barbieri:2003pr,Davoudiasl:2003me,Burdman:2003ya,Cacciapaglia:2004jz,Davoudiasl:2004pw,Barbieri:2004qk,Evans:2004rc,Perelstein:2004sc}. 
Our expressions for these zero-momentum parameters depend
only on the spectrum of heavy vector-boson masses; therefore, the minimum size of the deviations
present in these models is related to the upper bound on the heavy vector-boson masses
derived from unitarity. We find that these models are disfavored by precision electroweak 
data, independent of any assumptions about the background metric or the behavior of the
bulk coupling.

\section{Parameterizing Deviations from the Standard Model}

Barbieri {\it et al.} \cite{Barbieri:2004qk} choose parameters to describe four-fermion
electroweak processes using the transverse gauge-boson 
polarization amplitudes. Formally, all such processes can be
summarized in momentum space (at tree-level in the electroweak interactions, having integrated
out all heavy states, and ignoring external fermion masses) by the charged
current Lagrangian
\begin{equation}
{\cal L}_{cc}= {1\over 2} \left[\Pi_{W^+ W^-}(Q^2)\right]^{-1} J^\mu_+ J_{-\mu}~,
\end{equation}
and the neutral current Lagrangian
\begin{equation}
{\cal L}_{nc} = {1\over 2} 
\left(
\begin{array}{cc}
J_{3\mu} & J_{B\mu}
\end{array}
\right)
\left[
\begin{array}{cc}
{\Pi}_{W^3 W^3}(Q^2) & {\Pi}_{W^3 B}(Q^2) \\
{\Pi}_{W^3 B}(Q^2) & {\Pi}_{BB}(Q^2)
\end{array}
\right]^{-1}
\left(
\begin{array}{c}
J^\mu_{3} \\
J_B^\mu
\end{array}
\right)~,
\end{equation}
where the $\vec{J}^\mu$ and $J_B^\mu$ are the weak isospin and hypercharge
fermion currents respectively. All two-point correlation functions of fermionic currents
-- and therefore all four-fermion scattering amplitudes at tree-level -- 
can be read off from the appropriate element(s) of the inverse
gauge-boson polarization matrix. Throughout this paper, in order to
be consistent with \cite{Chivukula:2004pk}, we use $Q^2=-q^2$ to denote the
Euclidean momentum-squared.

Barbieri {\it et al.} proceed by defining the (approximate) electroweak couplings
\begin{equation}
{1\over g^2} \equiv \left[d\Pi_{W^+ W^-}(Q^2) \over d(-Q^2)\right]_{Q^2=0}
\ \ \ \ , \ \ \ 
{1\over {g'}^2} \equiv \left[{d\Pi_{BB}(Q^2)\over d(-Q^2)}\right]_{Q^2=0}~,
\label{defgs}
\end{equation}
and the electroweak scale\footnote{Our definition of $v$ differs from that used in ref.
\protect\cite{Barbieri:2004qk} by $\sqrt{2}$.}
\begin{equation}
v^2 \equiv -4\,\Pi_{W^+ W^-}(0)  = (\sqrt{2} G_F)^{-1}\approx (246\,{\rm GeV})^2~.
\end{equation}
In terms of the polarization functions and these constants, the authors of \cite{Barbieri:2004qk}
define the parameters
\begin{eqnarray}
\hat{S} & \equiv & g^2 \left[{d\Pi_{W^3 B} (Q^2)\over d(-Q^2)}\right]_{Q^2=0}~, \label{defshat}\\
\hat{T} & \equiv &  {g^2 \over M^2_W}\left(\Pi_{W^3W^3}(0)-\Pi_{W^+ W^-}(0)\right)~,\label{defthat}\\
W & \equiv & {g^2 M^2_W \over 2} \left[{d^2 \Pi_{W^3 W^3} (Q^2)\over d(-Q^2)^2}\right]_{Q^2=0}~,\label{defw}\\
Y & \equiv & {{g'}^2 M^2_W \over 2} \left[{d^2 \Pi_{BB}(Q^2) \over d(-Q^2)^2}\right]_{Q^2=0}~. \label{defy}
\end{eqnarray}
In any non-standard electroweak model in which all of the relevant effects occur 
{\it only} in the correlation function
of fermionic electroweak gauge currents,\footnote{And not, for example, through extra gauge-bosons or
compositeness operators involving the $B-L$ or weak isosinglet currents \protect\cite{Chivukula:1987zq}.}
the values of these four parameters \cite{Barbieri:2004qk} summarize the leading deviations 
in all four-fermion processes from the standard model predictions.  The quantities $\hat{U}$, $\hat{V}$ and $X$ defined in \cite{Barbieri:2004qk} describe higher-order effects.

While the parameters of eqns. (\ref{defshat})-(\ref{defy}) succinctly summarize the deviations
from the standard model in any universal extension, they do not
correspond simply to the on-shell properties of the $Z$ or $W$ bosons, the properties most
easily calculated when considering models with extra gauge bosons, for example. Instead, it
is useful to characterize\footnote{The matrix element definitions that follow are slight 
generalizations of those
proposed in \protect\cite{Chivukula:2004pk}. The ones proposed here allow for the low-energy
$\rho$-parameter to deviate from one and, consistent with the arguments of 
\protect\cite{Barbieri:2004qk}, have $U=0$. We have also changed the overall sign
of the matrix elements to conform to the usual definitions, and have used the
relation $M^2_Z \approx  \pi\alpha/(\sqrt{2} G_F s^2 c^2)$ to simplify the coefficient of $T$.} 
the matrix element for four-fermion neutral weak  current processes by
\begin{eqnarray}
-{\cal M}_{NC} = e^2 {{\cal Q}{\cal Q}' \over Q^2} 
& + &
\dfrac{(I_3-s^2 {\cal Q}) (I'_3 - s^2 {\cal Q}')}
	{\left({s^2c^2 \over e^2}-{S\over 16\pi}\right)Q^2 +
		{1\over 4 \sqrt{2} G_F}\left(1-\alpha T +{\alpha \delta \over 4 s^2 c^2}\right)
		} 
\label{eq:NC4} \\ \nonumber & \ \ & \\
&+&
\sqrt{2} G_F \,{\alpha \delta\over  s^2 c^2}\, I_3 I'_3 
+ 4 \sqrt{2} G_F  \left( \Delta \rho - \alpha T\right)({\cal Q}-I_3)({\cal Q}'-I_3')~,
\nonumber 
\end{eqnarray}
and the matrix element for charged currents by 
\begin{eqnarray}
  - {\cal M}_{\rm CC}
  =  \dfrac{(I_{+} I'_{-} + I_{-} I'_{+})/2}
             {\left(\dfrac{s^2}{e^2}-\dfrac{S}{16\pi}\right)Q^2
             +{1\over 4 \sqrt{2} G_F}\left(1+{\alpha \delta \over 4 s^2 c^2}\right)
            }
        + \sqrt{2} G_F\, {\alpha  \delta\over s^2 c^2} \, {(I_{+} I'_{-} + I_{-} I'_{+}) \over 2}~.
\label{eq:CC3}
\end{eqnarray}
Here $I^{(\prime)}_a$ and ${\cal Q}^{(\prime)}$ are weak isospin and charge
of the corresponding fermion, $\alpha = e^2/4\pi$, $G_F$ is the usual Fermi constant,
and the weak mixing angle (as defined by the on-shell $Z$ coupling) is denoted by $s^2$ 
($c^2\equiv 1-s^2$).

Some comments about the amplitudes in eqns. (\ref{eq:NC4}) and (\ref{eq:CC3}) are in
order. First, $\Delta \rho$ corresponds to the deviation from unity of the ratio of the strengths of
low-energy isotriplet weak neutral-current scattering and charged-current scattering.
$S$ and $T$ are the familiar oblique electroweak parameters \cite{Peskin:1992sw,Altarelli:1990zd,Altarelli:1991fk}, 
as determined by examining the {\it on-shell} properties of the $Z$ and $W$ bosons.
Finally, the contact interactions proportional to $\alpha \delta$ and ($\Delta \rho - \alpha T$)
correspond to ``universal non-oblique'' corrections. They are ``universal'' in the sense that
they can be seen to arise from effective operators proportional to
$\vec{J}^2_\mu$ and $J^2_B$, and therefore
modify the correlation function of fermionic electroweak currents. They are ``non-oblique''
in the sense that they do not correspond to deviations of the on-shell $W$- and $Z$-boson
propagators. As shown in \cite{Chivukula:2004pk}, such universal non-oblique effects occur
in a variety of Higgsless models of electroweak symmetry breaking -- and 
the presence of such effects need not lead to deviations in the low-energy $\rho$ parameter
from one.

Relating the parameters $\alpha S$, $\alpha T$, $\alpha \delta$, and $\Delta \rho$ to
$\hat{S}$, $\hat{T}$, $W$, and $Y$ is straightforward\footnote{An alternative procedure
is based on interpreting the matrix elements of eqns. (\protect\ref{eq:NC4}) and (\protect\ref{eq:CC3})
in terms of effective operators and relating them, using the equations of motion
\protect\cite{Strumia:1999jm}, to the operator analysis presented in 
\protect\cite{Barbieri:2004qk}.}: inverting the charged-current
matrix element of eqn. (\ref{eq:CC3}) yields $\Pi_{W^+ W^-}(Q^2)$, and finding the
inverse of the $2 \times 2$ matrix in the space of currents $(J_{3\mu},J_{B\mu})$, defined
implicitly by eqn. (\ref{eq:NC4}), yields the neutral-current matrix ${\bf \Pi}(Q^2)$.
In the limit where all corrections to the standard model go to zero, one finds
\begin{equation}
\Pi^{SM}_{W^+ W^-}(Q^2) = - {s^2\over e^2}\left[
Q^2 + {e^2\over 4\sqrt{2}s^2 G_F}\right]~,
\end{equation}
and 

\begin{equation}
{\bf \Pi}^{SM}(Q^2) = -{1\over e^2}\,\left(
\begin{array}{cc}
s^2 Q^2+c^2 s^2 \mu^2_Z & -s^2 c^2 \mu^2_Z \\
-s^2 c^2 \mu^2_Z & c^2 Q^2+c^2 s^2 \mu^2_Z
\end{array}
\right)~,
\label{pism}
\end{equation}
where we have defined
\begin{equation}
\mu^2_Z \equiv {e^2\over 4\sqrt{2}G_F s^2 c^2}~,
\label{eq:sdef}
\end{equation}
for convenience.
From these lowest-order expressions, we immediately find from eqn. (\ref{defgs})
\begin{equation}
{1\over g^2} \approx {e^2 \over s^2} \ \ \ \ , \ \ \ {1\over {g'}^2} \approx {e^2\over c^2}~,
\end{equation}
as expected.

Calculating ${\bf \Pi}(Q^2)$ to leading order in the deviations from the standard model,
one finds the relations\footnote{Note that, although $s^2$ is defined implicitly 
in eqn. (\protect\ref{eq:NC4}) in terms of the
on-shell $Z$-boson couplings, to this order in (small) deviations from the standard model,
any definition of the weak mixing angle can be used consistently in (\ref{eq:def-a}) - (\ref{eq:def-d}).}
\begin{eqnarray}
\hat{S} & = & {1\over 4s^2}\left(\alpha S + 4 c^2 (\Delta \rho - \alpha T) + {\alpha \delta \over c^2}\right) \label{eq:def-a}\\
\hat{T} & = & \Delta \rho \\
W & = & {\alpha \delta \over 4 s^2 c^2} \\
Y & = & {c^2 \over s^2} (\Delta \rho - \alpha T)~.\label{eq:def-d}
\end{eqnarray}
Inverting these relationships, we find
\begin{eqnarray}
\alpha S & = & 4 s^2 ( \hat{S} - Y - W) \\
\alpha T & = & \hat{T}-{s^2 \over c^2} Y \\
\alpha \delta & = & 4 s^2 c^2 W \\
\Delta \rho & = & \hat{T}~.
\end{eqnarray}

In the absence of any universal non-oblique corrections, $W=Y=0$, one
finds the relations
\begin{equation}
\hat{S} = {\alpha  S \over 4 s^2} \ \ \ , \ \ \hat{T} = \alpha T~,
\end{equation}
given in \cite{Barbieri:2004qk}. 

Note that it is the non-oblique universal corrections described by $Y$ that mark the difference between $\Delta \rho$ and $T$.  In a model with $Y=0$ we have $\Delta\rho = \hat{T} = \alpha T$, so that the case of greatest phenomenological interest with  $\Delta \rho = 0$ would also have vanishing $\hat{T}$ and $T$.  However, in a model with non-zero $Y$, focusing on the case with $\Delta\rho = 0$ ensures that $\hat{T}$ vanishes, but still allows $\alpha T$ to be non-zero.  

\section{Application to Higgsless Models}

\begin{figure}[ht]
\centering
\includegraphics[width=15cm]{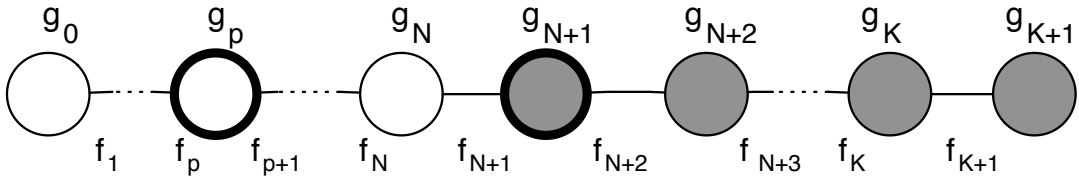}
\caption{
Moose diagram for the class of models discussed in this letter. All $N+1$ of the 
$SU(2)$ gauge groups are shown as open
circles;  all $M+1$ of the $U(1)$ gauge groups, as shaded circles; and $K=N+M$.
The fermions couple to gauge
gauge groups $p$ and $N+1$.  The values of the gauge couplings $g_i$ and decay constants $f_i$ are arbitrary.}
\label{Fig:TheMoose}
\end{figure}

We may now apply these results to Higgsless models. 
Using deconstruction \cite{Arkani-Hamed:2001ca,Hill:2000mu} , 
the most general Higgsless model in which the low-energy
$\rho$ parameter is one \cite{Chivukula2004} is shown diagrammatically in 
Fig. \ref{Fig:TheMoose} (in ``moose notation'' 
\cite{Georgi:1986hf,Arkani-Hamed:2001ca}).\footnote{These models generalize 
those considered in
\protect\cite{Chivukula:2004pk}, by allowing for fermion couplings to an arbitrary
$SU(2)$ group along the moose.}
 These models incorporate an
$SU(2)^{N+1} \times U(1)^{M+1}$ gauge group, and $N+1$ 
nonlinear $(SU(2)\times SU(2))/SU(2)$ sigma models adjacent to $M$
$(U(1) \times U(1))/U(1)$ sigma models in which the global symmetry groups 
in adjacent sigma models are identified with the corresponding factors of the gauge group.
The Lagrangian for this model at $O(p^2)$ is given by
\begin{equation}
  {\cal L}_2 =
  \frac{1}{4} \sum_{j=1}^{N+M+1} f_j^2 \mbox{tr}\left(
    (D_\mu U_j)^\dagger (D^\mu U_j) \right)
  - \sum_{j=0}^{N+M+1} \dfrac{1}{2g_j^2} \mbox{tr}\left(
    F^j_{\mu\nu} F^{j\mu\nu}
    \right),
\label{lagrangian}
\end{equation}
with
\begin{equation}
  D_\mu U_j = \partial_\mu U_j - i A^{j-1}_\mu U_j 
                               + i U_j A^{j}_\mu,
\end{equation}
where all  gauge fields $A^j_\mu$ $(j=0,1,2,\cdots, N+M+1)$ are dynamical. The first
$N+1$ gauge fields ($j=0,1,\ldots, N$) correspond to $SU(2)$ gauge groups; the other $M+1$ gauge
fields ($j=N+1, N+2, \ldots, N+M+1$) correspond to $U(1)$ gauge groups.  The symmetry breaking between
the $A^{N}_\mu$ and $A^{N+1}_\mu$ follows an $SU(2)_L \times SU(2)_R/SU(2)_V$ symmetry
breaking pattern with the $U(1)$ embedded as the $T_3$-generator of $SU(2)_R$.

The fermions in this model take their weak interactions from the $SU(2)$ group at $j=p$ and their hypercharge interactions from the $U(1)$ group with $j=N+1$, at the interface between the $SU(2)$ and $U(1)$ groups \footnote{As discussed in \cite{Chivukula2004}, the choice to associate this $U(1)$ group with the fermions' hypercharge is what guarantees that $\rho$ will equal one.}  The neutral current couplings to the fermions are thus written as
\begin{equation}
J^\mu_3 A^p_\mu + J^\mu_B A^{N+1}_\mu~,
\label{eq:current}
\end{equation}
while the charged current couplings arise from
\begin{equation}
{1 \over \sqrt{2}} J^\mu_{\pm} A^{p\mp}_\mu~.
\end{equation}

Generalizing the calculations of \cite{Chivukula:2004pk} one may calculate the
polarization functions $\Pi_{W^+ W^-}(Q^2)$ and ${\bf \Pi}(Q^2)$ at tree-level
\cite{Chivukula2004}. We find that
\begin{equation}
\Pi_{W^+ W^-}(Q^2) = \Pi_{W^3 W^3}(Q^2)~,
\end{equation}
and therefore the parameter $\hat{T}$, as well as the higher-order
parameters $\hat{U}$ and $\hat{V}$ \cite{Barbieri:2004qk},  vanishes identically in any of these
models. 

The results for the non-zero parameters are most conveniently expressed in terms
of the eigenvalues of various sub-matrices of the full neutral vector-boson mass-squared matrix.
Generalizing the usual mathematical
notation for ``open'' and ``closed'' intervals, we may denote the
neutral-boson mass matrix $M^2_Z$ as $M^2_{[0,N+M+1]}$ --- {\it i.e.}
it is the mass matrix for the entire moose running from site $0$ to site $N+M+1$ including
the gauge couplings of both endpoint groups. Analogously, the charged-boson mass matrix $M^2_W$ is
$M^2_{[0,N+1)}$ --- it is the mass matrix for the moose running from site $0$ to link
$N+1$, but not including the gauge couping at site $N+1$.
Using this notation, we define sub-matrices:
\begin{eqnarray}
{\cal M}^2_p & = & M^2_{[0,p)} \\
{\cal  M}^2_r & = & M^2_{(p,N+1)} \\
{\cal  M}^2_q & = & M^2_{(N+1,N+M+1]} 
\end{eqnarray}
of the neutral gauge-boson mass-squared matrix.  They fit together inside $M_Z^2$ as follows:
\begin{equation}
{\tiny
  M_{Z}^2 = \left(
    \begin{array}{cc|c|ccc|c|cc}
   & {\cal M}_p^2 & & & & & & & \\
   & &  - g^{}_{p-1} g^{}_{p} f_{p}^2/4 & & & & & & \\
   \hline
   &  - g^{}_{p-1} g^{}_{p} f_{p+1}^2/4
      &   g_{p}^2 (f_{p}^2 + f_{p+1}^2)/4
      & - g^{}_{p} g^{}_{p+1} f_{p+1}^2/4
      & & & & & \\
      \hline
    & &  - g^{}_{p} g^{}_{p+1} f_{p+1}^2/4 & & & & & & \\
    & & & & {\cal M}_{r}^2  & & & & \\
    & & & & & & - g^{}_{N} g^{}_{N+1} f_{N+1}^2/4 & &\\
    \hline
    & & & & & - g^{}_{N} g^{}_{N+1} f_{N+2}^2/4 
      &   g_{N+1}^2 (f_{N+1}^2 + f_{N+2}^2)/4
      & - g^{}_{N+1} g^{}_{N+2} f_{N+2}^2/4
      & \\
      \hline
      & & & & & & - g^{}_{N+1} g^{}_{N+1} f_{N+2}^2/4 & & \\
      & & & & & & &  {\cal{M}}_{q}^2 \\
    \end{array}
  \right).
}
\label{eq:mass_matrixN3}
\end{equation}

In the phenomenologically relevant limit, in which the only light vector bosons
correspond to the usual $\gamma$, $W$, and $Z$, the eigenvalues of these matrices
(${\mathsf m}^2_{\hat{p},\hat{r},\hat{q}}$ respectively) must be large, 
${\mathsf m}^2_{\hat{p},\hat{r},\hat{q}} \gg M^2_{W,Z}$ \cite{Chivukula2004}.
It is therefore appropriate to expand in inverse powers of the large mass eigenvalues.
We define 
\begin{equation}
\Sigma_Z = \sum_{\hat{z} = 1}^{N+M} {1\over {\mathsf m}^2_{\hat{z}}}\ \ \ \ , \ \ \ 
\Sigma_W = \sum_{\hat{w}=1}^{N} {1\over {\mathsf m}^2_{\hat{w}}}~,
\end{equation}
where the sums run only over the heavy eigenstates ({\it i.e.} they exclude the
light $W$, light $Z$ and photon), and
\begin{equation}
\Sigma_p = \sum_{\hat{p}=0}^{p-1} {1\over {\mathsf m}^2_{\hat{p}}}\ \ ,\ \ 
\Sigma_r=\sum_{\hat{r}=p+1}^{N}{1\over {\mathsf m}^2_{\hat{r}} }\ \ , \ \ 
\Sigma_q = \sum_{\hat{q}=N+2}^{N+M+1} {1\over {\mathsf m}^2_{\hat{q}}}~.
\end{equation}
where the sums run over all of the submatrix eigenvalues.  

We can write the electroweak parameters in terms of these sums over eigenvalues.  The on-shell parameters (recalling that $\Delta\rho \equiv 0$) take the form  \cite{Chivukula2004}
\begin{eqnarray}
\alpha S & = & 4 s^2 M^2_W(\Sigma_Z-\Sigma_p - \Sigma_q) \\
\alpha T & = & s^2 M^2_Z(\Sigma_Z - \Sigma_W - \Sigma_q) \\
{\alpha \delta \over c^2} & = & -4 s^2 M^2_W(\Sigma_W - \Sigma_p - \Sigma_r) ~.
\end{eqnarray}
Clearly it is possible for $S$ to be small or even negative.  In the case where the fermions couple to the $SU(2)$ group at the left end of the moose (i.e., $p=0$), these reduce to the expressions found in \cite{Chivukula:2004pk}:   $\Sigma_p = 0$, our $\Sigma_q$ is equivalent to $\Sigma_M$ in the earlier paper, and $M_W^2$ may be used in place of $c^2 M_Z^2$ to leading order in these expressions.  

Using the relations (\ref{eq:def-a}) - (\ref{eq:def-d}) we can write the zero-momentum parameters as 
\begin{eqnarray}
\hat{S} & = & M^2_W\, \Sigma_r  > 0\\
\hat{T} & = & 0 \\
W & = & - M^2_W ( \Sigma_W - \Sigma_p - \Sigma_r) \\
Y & = & - M^2_W(\Sigma_Z - \Sigma_W - \Sigma_q)~,
\end{eqnarray}
to leading non-trivial order. 
As noted, the parameter $\hat{S}$ is always strictly positive, in agreement with 
the arguments presented in \cite{Barbieri:2003pr}. Re-expressing $\hat{S}$ in
terms of the on-shell parameters (setting $\Delta \rho =0$ as appropriate
in this class of models) we see that
\begin{equation}
\hat{S}={\alpha \over 4s^2}\left(S - 4 c^2 T + {\delta \over c^2}\right)= M^2_W \Sigma_r > 0~,
\label{eq:summary}
\end{equation}
generalizing the result of \cite{Chivukula:2004pk}.
Furthermore, we see that, due to the presence of non-oblique universal corrections  the positivity of $\hat{S}$ is not in contradiction with small, or even negative, values of
$\alpha S$ \cite{Cacciapaglia:2004jz,Davoudiasl:2004pw}.

\begin{figure}[ht]
\centering
\includegraphics[width=10cm]{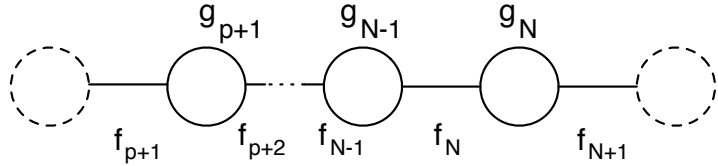}
\caption{
Moose diagram whose vector boson mass matrix corresponds to
${\cal M}^2_r$, which runs from link $p+1$ to link $N+1$  of the original moose in Fig. 1. 
The endpoints are
ungauged and the dashed circles indicate global
groups. The $F$-constant of the
physical pions of this model is $v$.
 }
\label{Fig:TherMoose}
\end{figure}

In any unitary theory \cite{SekharChivukula:2001hz, Chivukula:2002ej}, we expect the mass of the lightest additional 
vector to be less than $\sqrt{8 \pi} v$ ($v\approx 246$ GeV), the scale at
which $WW$ spin-0 isospin-0 elastic scattering would violate unitarity in the 
standard model in the absence of a higgs
boson \cite{Dicus:1973vj,Cornwall:1973tb, Cornwall:1974km,Lee:1977yc, Lee:1977eg,Veltman:1977rt}. In the case that $M^2_W\, \Sigma_{Z,W,p,q,r} \ll 1$, the Goldstone boson corresponding to
the longitudinal $W$ is approximately the pion of the model shown in Fig. 2
\cite{Chivukula2004}. Unitarity, therefore, requires that the lightest 
eigenvalue of the matrix ${\cal M}^2_r$ must be
of order $8\pi v^2$ or lighter.
Evaluating eqn. (\ref{eq:summary}) then reveals 
$S-4 c^2 T + \delta/c^2$ to be of order one-half  or larger, 
generalizing the result of \cite{Chivukula:2004pk}. The corresponding value of $\hat{S}$ this large is
disfavored by precision electroweak data \cite{Barbieri:2004qk}.\footnote{In a recent paper
\cite{Perelstein:2004sc}, Perelstein has argued that the higher-order corrections expected
to be present in any QCD-like ``high-energy'' completion of a Higgsless theory are
also likely to be large. We have calculated the tree-level corrections expected
independent of the form of the high-energy completion.}

\section{Summary}

In this letter about universal theories, we have related the parameters ($\hat{S}$, $\hat{T}$, $W$, $Y$,) introduced by Barbieri {\it et al.}
to describe zero-momentum deviations of the electroweak interactions from their standard model forms 
to the parameters ($S$, $T$, $\delta$, $\Delta \rho$) calculated in terms of the on-shell properties
of the $W$ and $Z$ bosons. We have presented the results of a calculation \cite{Chivukula2004} 
of these parameters in the most general Higgsless model in which the low-energy $\rho$ parameter
is one. Our results demonstrate the importance of the universal non-oblique corrections
which are generally present in these models. These results also
elucidate the relationship between the various calculations of precision electroweak
parameters in Higgsless models.  

Specifically, we find $\hat{S} = (\alpha/4s^2)(S - 4 c^2 T + \delta/c^2) > 0$ which agrees with and extends previous findings \cite{Chivukula:2004pk,Barbieri:2004qk,Barbieri:2003pr}.  Moreover, we find that unitarity considerations constrain $4 s^2 \hat{S}/\alpha$ to be greater than or of order one-half in these models, a value so large as to be severely disfavored \cite{Barbieri:2004qk} by precision electroweak data.

The details of the calculations of the various correlation
functions in Higgsless models, the generalization to models
with $\Delta \rho \neq 0$, and the connection to an expansion in large ``bulk''
coupling \cite{Georgi:2004iy}, will be presented in \cite{Chivukula2004}.

\section{Note Added in Proof}

After the submission of this manuscript, a new class of Higgsless
models with delocalized fermions has been proposed 
\cite{Cacciapaglia:2004rb,Foadi:2004ps}, and it
has been shown that the delocalization of the fermions can be adjusted
to minimize the deviations of the electroweak interactions from their
Standard Model forms. The techniques discussed here and in \cite{Chivukula2004} 
must be extended to accommodate fermion delocalization, and this topic is
under current investigation.

 \acknowledgments

We would like to thank Nick Evans, Howard Georgi, and Carl Schmidt  for discussions. 
R.S.C. and E.H.S. acknowledge the
hospitality of the Aspen Center for Physics where some of this work was completed.
M.K. acknowledges support by the 21st Century COE Program of Nagoya University 
provided by JSPS (15COEG01). M.T.'s work is supported in part by the JSPS Grant-in-Aid for Scientific Research No.16540226. H.J.H. is supported by the US Department of Energy grant
DE-FG03-93ER40757.

%%%%%%%%%%%%%%%%%%%%%%%%%%%%%%%%%%%%%%%%%%%%%%%%%%%%%%

%\bibliography{universal.bib}

\begin{thebibliography}{99}
\bibitem{Csaki:2003dt}
C.~Csaki, C.~Grojean, H.~Murayama, L.~Pilo, and J.~Terning, {\it Gauge theories
  on an interval: Unitarity without a higgs},
  \href{http://xxx.lanl.gov/abs/hep-ph/0305237}{{\tt hep-ph/0305237}}.

\bibitem{SekharChivukula:2001hz}
R.~Sekhar~Chivukula, D.~A. Dicus, and H.-J. He, {\it Unitarity of compactified
  five dimensional yang-mills theory},  {\em Phys. Lett.} {\bf B525} (2002)
  175--182, [\href{http://xxx.lanl.gov/abs/hep-ph/0111016}{{\tt
  hep-ph/0111016}}].

\bibitem{Chivukula:2002ej}
R.~S. Chivukula and H.-J. He, {\it Unitarity of deconstructed five-dimensional
  yang-mills theory},  {\em Phys. Lett.} {\bf B532} (2002) 121--128,
  [\href{http://xxx.lanl.gov/abs/hep-ph/0201164}{{\tt hep-ph/0201164}}].

\bibitem{Chivukula:2003kq}
R.~S. Chivukula, D.~A. Dicus, H.-J. He, and S.~Nandi, {\it Unitarity of the
  higher dimensional standard model},  {\em Phys. Lett.} {\bf B562} (2003)
  109--117, [\href{http://xxx.lanl.gov/abs/hep-ph/0302263}{{\tt
  hep-ph/0302263}}].

\bibitem{Higgs:1964ia}
P.~W. Higgs, {\it Broken symmetries, massless particles and gauge fields},
  {\em Phys. Lett.} {\bf 12} (1964) 132--133.

\bibitem{Maldacena:1998re}
J.~M. Maldacena, {\it The large n limit of superconformal field theories and
  supergravity},  {\em Adv. Theor. Math. Phys.} {\bf 2} (1998) 231--252,
  [\href{http://xxx.lanl.gov/abs/hep-th/9711200}{{\tt hep-th/9711200}}].

\bibitem{Gubser:1998bc}
S.~S. Gubser, I.~R. Klebanov, and A.~M. Polyakov, {\it Gauge theory correlators
  from non-critical string theory},  {\em Phys. Lett.} {\bf B428} (1998)
  105--114, [\href{http://xxx.lanl.gov/abs/hep-th/9802109}{{\tt
  hep-th/9802109}}].

\bibitem{Witten:1998qj}
E.~Witten, {\it Anti-de sitter space and holography},  {\em Adv. Theor. Math.
  Phys.} {\bf 2} (1998) 253--291,
  [\href{http://xxx.lanl.gov/abs/hep-th/9802150}{{\tt hep-th/9802150}}].

\bibitem{Aharony:1999ti}
O.~Aharony, S.~S. Gubser, J.~M. Maldacena, H.~Ooguri, and Y.~Oz, {\it Large n
  field theories, string theory and gravity},  {\em Phys. Rept.} {\bf 323}
  (2000) 183--386, [\href{http://xxx.lanl.gov/abs/hep-th/9905111}{{\tt
  hep-th/9905111}}].

\bibitem{Weinberg:1979bn}
S.~Weinberg, {\it Implications of dynamical symmetry breaking: An addendum},
  {\em Phys. Rev.} {\bf D19} (1979) 1277--1280.

\bibitem{Susskind:1979ms}
L.~Susskind, {\it Dynamics of spontaneous symmetry breaking in the weinberg-
  salam theory},  {\em Phys. Rev.} {\bf D20} (1979) 2619--2625.

\bibitem{Holdom:1981rm}
B.~Holdom, {\it Raising the sideways scale},  {\em Phys. Rev.} {\bf D24} (1981)
  1441.

\bibitem{Holdom:1985sk}
B.~Holdom, {\it Techniodor},  {\em Phys. Lett.} {\bf B150} (1985) 301.

\bibitem{Yamawaki:1986zg}
K.~Yamawaki, M.~Bando, and K.-i. Matumoto, {\it Scale invariant technicolor
  model and a technidilaton},  {\em Phys. Rev. Lett.} {\bf 56} (1986) 1335.

\bibitem{Appelquist:1986an}
T.~W. Appelquist, D.~Karabali, and L.~C.~R. Wijewardhana, {\it Chiral
  hierarchies and the flavor changing neutral current problem in technicolor},
  {\em Phys. Rev. Lett.} {\bf 57} (1986) 957.

\bibitem{Appelquist:1987tr}
T.~Appelquist and L.~C.~R. Wijewardhana, {\it Chiral hierarchies and chiral
  perturbations in technicolor},  {\em Phys. Rev.} {\bf D35} (1987) 774.

\bibitem{Appelquist:1987fc}
T.~Appelquist and L.~C.~R. Wijewardhana, {\it Chiral hierarchies from slowly
  running couplings in technicolor theories},  {\em Phys. Rev.} {\bf D36}
  (1987) 568.

\bibitem{Barbieri:2004qk}
R.~Barbieri, A.~Pomarol, R.~Rattazzi, and A.~Strumia, {\it Electroweak symmetry
  breaking after lep1 and lep2},
  \href{http://xxx.lanl.gov/abs/hep-ph/0405040}{{\tt hep-ph/0405040}}.

\bibitem{Arkani-Hamed:2001ca}
N.~Arkani-Hamed, A.~G. Cohen, and H.~Georgi, {\it (de)constructing dimensions},
   {\em Phys. Rev. Lett.} {\bf 86} (2001) 4757--4761,
  [\href{http://xxx.lanl.gov/abs/hep-th/0104005}{{\tt hep-th/0104005}}].

\bibitem{Hill:2000mu}
C.~T. Hill, S.~Pokorski, and J.~Wang, {\it Gauge invariant effective lagrangian
  for kaluza-klein modes},  {\em Phys. Rev.} {\bf D64} (2001) 105005,
  [\href{http://xxx.lanl.gov/abs/hep-th/0104035}{{\tt hep-th/0104035}}].

\bibitem{Chivukula:2004pk}
R.~S. Chivukula, E.~H. Simmons, H.-J. He, M.~Kurachi, and M.~Tanabashi, {\it
  The structure of corrections to electroweak interactions in higgsless
  models},  \href{http://xxx.lanl.gov/abs/hep-ph/0406077}{{\tt
  hep-ph/0406077}}.

\bibitem{Chivukula2004}
R.~S. Chivukula, E.~H. Simmons, H.-J. He, M.~Kurachi, and M.~Tanabashi, 
``Electroweak corrections and unitarity in linear moose models,''
hep-ph/0410154.

\bibitem{Csaki:2003zu}
C.~Csaki, C.~Grojean, L.~Pilo, and J.~Terning, {\it Towards a realistic model
  of higgsless electroweak symmetry breaking},  {\em Phys. Rev. Lett.} {\bf 92}
  (2004) 101802, [\href{http://xxx.lanl.gov/abs/hep-ph/0308038}{{\tt
  hep-ph/0308038}}].

\bibitem{Nomura:2003du}
Y.~Nomura, {\it Higgsless theory of electroweak symmetry breaking from warped
  space},  {\em JHEP} {\bf 11} (2003) 050,
  [\href{http://xxx.lanl.gov/abs/hep-ph/0309189}{{\tt hep-ph/0309189}}].

\bibitem{Barbieri:2003pr}
R.~Barbieri, A.~Pomarol, and R.~Rattazzi, {\it Weakly coupled higgsless
  theories and precision electroweak tests},
  \href{http://xxx.lanl.gov/abs/hep-ph/0310285}{{\tt hep-ph/0310285}}.

\bibitem{Davoudiasl:2003me}
H.~Davoudiasl, J.~L. Hewett, B.~Lillie, and T.~G. Rizzo, {\it Higgsless
  electroweak symmetry breaking in warped backgrounds: Constraints and
  signatures},  \href{http://xxx.lanl.gov/abs/hep-ph/0312193}{{\tt
  hep-ph/0312193}}.

\bibitem{Burdman:2003ya}
G.~Burdman and Y.~Nomura, {\it Holographic theories of electroweak symmetry
  breaking without a higgs boson},
  \href{http://xxx.lanl.gov/abs/hep-ph/0312247}{{\tt hep-ph/0312247}}.

\bibitem{Cacciapaglia:2004jz}
G.~Cacciapaglia, C.~Csaki, C.~Grojean, and J.~Terning, {\it Oblique corrections
  from higgsless models in warped space},
  \href{http://xxx.lanl.gov/abs/hep-ph/0401160}{{\tt hep-ph/0401160}}.

\bibitem{Davoudiasl:2004pw}
H.~Davoudiasl, J.~L. Hewett, B.~Lillie, and T.~G. Rizzo, {\it Warped higgsless
  models with ir-brane kinetic terms},  {\em JHEP} {\bf 05} (2004) 015,
  [\href{http://xxx.lanl.gov/abs/hep-ph/0403300}{{\tt hep-ph/0403300}}].

\bibitem{Evans:2004rc}
N.~Evans and P.~Membry, {\it Higgless w unitarity from decoupling
  deconstruction},  \href{http://xxx.lanl.gov/abs/hep-ph/0406285}{{\tt
  hep-ph/0406285}}.

\bibitem{Perelstein:2004sc}
M.~Perelstein, {\it Gauge-assisted technicolor?},
  \href{http://xxx.lanl.gov/abs/hep-ph/0408072}{{\tt hep-ph/0408072}}.

\bibitem{Chivukula:1987zq}
R.~S. Chivukula and H.~Georgi, {\it Phenomenology of composite technicolor
  standard models},  {\em Phys. Rev.} {\bf D36} (1987) 2102.

\bibitem{Peskin:1992sw}
M.~E. Peskin and T.~Takeuchi, {\it Estimation of oblique electroweak
  corrections},  {\em Phys. Rev.} {\bf D46} (1992) 381--409.

\bibitem{Altarelli:1990zd}
G.~Altarelli and R.~Barbieri, {\it Vacuum polarization effects of new physics
  on electroweak processes},  {\em Phys. Lett.} {\bf B253} (1991) 161--167.

\bibitem{Altarelli:1991fk}
G.~Altarelli, R.~Barbieri, and S.~Jadach, {\it Toward a model independent
  analysis of electroweak data},  {\em Nucl. Phys.} {\bf B369} (1992) 3--32.

\bibitem{Strumia:1999jm}
A.~Strumia, {\it Bounds on kaluza-klein excitations of the sm vector bosons
  from electroweak tests},  {\em Phys. Lett.} {\bf B466} (1999) 107--114,
  [\href{http://xxx.lanl.gov/abs/hep-ph/9906266}{{\tt hep-ph/9906266}}].

\bibitem{Georgi:1986hf}
H.~Georgi, {\it A tool kit for builders of composite models},  {\em Nucl.
  Phys.} {\bf B266} (1986) 274.

\bibitem{Dicus:1973vj}
D.~A. Dicus and V.~S. Mathur, {\it Upper bounds on the values of masses in
  unified gauge theories},  {\em Phys. Rev.} {\bf D7} (1973) 3111--3114.

\bibitem{Cornwall:1973tb}
J.~M. Cornwall, D.~N. Levin, and G.~Tiktopoulos, {\it Uniqueness of
  spontaneously broken gauge theories},  {\em Phys. Rev. Lett.} {\bf 30} (1973)
  1268--1270.

\bibitem{Cornwall:1974km}
J.~M. Cornwall, D.~N. Levin, and G.~Tiktopoulos, {\it Derivation of gauge
  invariance from high-energy unitarity bounds on the s - matrix},  {\em Phys.
  Rev.} {\bf D10} (1974) 1145.

\bibitem{Lee:1977yc}
B.~W. Lee, C.~Quigg, and H.~B. Thacker, {\it The strength of weak interactions
  at very high-energies and the higgs boson mass},  {\em Phys. Rev. Lett.} {\bf
  38} (1977) 883.

\bibitem{Lee:1977eg}
B.~W. Lee, C.~Quigg, and H.~B. Thacker, {\it Weak interactions at very
  high-energies: The role of the higgs boson mass},  {\em Phys. Rev.} {\bf D16}
  (1977) 1519.

\bibitem{Veltman:1977rt}
M.~J.~G. Veltman, {\it Second threshold in weak interactions},  {\em Acta Phys.
  Polon.} {\bf B8} (1977) 475.

\bibitem{Georgi:2004iy}
H.~Georgi, {\it Fun with higgsless theories},
  \href{http://xxx.lanl.gov/abs/hep-ph/0408067}{{\tt hep-ph/0408067}}.

%\cite{Cacciapaglia:2004rb}
\bibitem{Cacciapaglia:2004rb}
G.~Cacciapaglia, C.~Csaki, C.~Grojean and J.~Terning,
``Curing the ills of Higgsless models: The S parameter and unitarity,"
hep-ph/0409126.
%%CITATION = HEP-PH 0409126;%%

%\cite{Foadi:2004ps}
\bibitem{Foadi:2004ps}
R.~Foadi, S.~Gopalakrishna and C.~Schmidt,
``Effects of fermion localization in Higgsless theories and electroweak
constraints,"  hep-ph/0409266.
%%CITATION = HEP-PH 0409266;%%



 

\end{thebibliography}
%\bibliographystyle{JHEP}

\end{document}